\documentclass{article}
\usepackage{arxiv}
\usepackage{nicefrac}       
\usepackage{microtype}      
\usepackage{booktabs}       
\usepackage[square,sort,comma,numbers]{natbib}
\usepackage{doi}
\usepackage{amsmath}
\usepackage{amssymb}
\usepackage{caption}
\usepackage{float}
\usepackage[T1]{fontenc}
\usepackage{graphicx}
\usepackage[utf8]{inputenc}
\usepackage{mathtools}
\usepackage{subcaption}
\usepackage{wasysym}
\begin{document}

\title{Phonon dynamic behaviors induced by amorphous interlayer at heterointerfaces}


\author{ 
        Quanjie Wang\\
        Institute of Micro/Nano Electromechanical System College of Mechanical Engineering\\
        State Key Laboratory for Modification of Chemical Fibers and Polymer Materials Donghua University\\
        Shanghai, 201620, China. \\
	\And
	Jie Zhang\\
	Institute of Artificial Intelligence\\
        Donghua University\\
        Shanghai, 201620, China. \\
        \And
        Vladimir Chernysh \\
	Department of Physical Electronics\\
        Lomonosov Moscow State University\\
        Russia. \\
        \And
        Xiangjun Liu\textsuperscript{*}\\
	Institute of Micro/Nano Electromechanical System College of Mechanical Engineering\\
        State Key Laboratory for Modification of Chemical Fibers and Polymer Materials Donghua University\\
        Shanghai, 201620, China. \\
        \texttt{xjliu@dhu.edu.cn} \\
}


\renewcommand{\shorttitle}{Phonon dynamic behaviors induced by amorphous interlayer at heterointerfaces}

\hypersetup{
pdftitle={Phonon dynamic behaviors induced by amorphous interlayer at heterointerfaces},
pdfsubject={q-bio.NC, q-bio.QM}, 
pdfauthor={Quanjie Wang, Jie Zhang, Vladimir Chernysh, Xiangjun Liu},
pdfkeywords={},
}

\maketitle

\begin{abstract}
Interface impedes heat flow in heterostructures and the interfacial thermal resistance (ITR) has become a critical issue for thermal dissipation in electronic devices. To explore the mechanism leading to the ITR, in this work, the dynamic behaviors of phonons passing through the GaN/AlN interface with an amorphous interlayer is investigated by using phonon wave packet simulation. It is found the amorphous interlayer significantly impedes phonon transport across the interface, and leads to remarkable phonon mode conversions, such as LA$\rightarrow$TA, TA$\rightarrow$LA, and LA$\rightarrow$TO conversion. However, due to mode conversion and inelastic scattering, we found a portion of high-frequency TA phonons, which are higher than the cut-off frequency and cannot transmit across the ideal sharp interface, can partially transmit across the amorphous interlayer, which introduces additional thermal transport channels through the interface and has positive effect on interfacial thermal conductance. According to phonon transmission coefficient, it is found the ITR increases with increasing of amorphous interlayer thickness L. The phonon transmission coefficient exhibits an obvious oscillation behavior, which is attributed to the multiple phonon scattering in the amorphous interlayer, and the oscillation period is further revealed to be consistent with the theoretical prediction by the two-beam interference equation. In addition, obvious phonon frequency shifts and phonon energy localization phenomena were
\end{abstract}

\section{Introduction}

Interface impedes heat flow in heterostructures and the interfacial thermal resistance (ITR) has become a critical issue for thermal dissipation and energy management in electronic devices, especially for high-speed and high-power electronic applications. For example, in GaN-based high-electron mobility transistors (HEMTs), it is predicted that 40$\%$ temperature rise in the channel layer is attributed to the high ITR between GaN and substrate, leading to the channel temperature over 100 $^{\circ}C$  \cite{lee2018thermal}. To alleviate this issue, various approaches with interface engineering have been developed to reduce the ITR, such as inserting buffer layer to mediate the vibrational mismatch  \cite{hu2011large} \cite{lee2017role},doping light/ isotope atoms to redistribute the phonon energy among different modes  \cite{lee2018thermal} \cite{li2019effect}, and enlarging interfacial contact area to increase the thermal transport channel  \cite{tao2017interlaced} \cite{lee2016nanostructures}.These approaches give rise to complex interfacial phonon scattering and transmission.  Deep insight into the interplay between interface microstructures and phonon dynamic behavior can provide scientific guidance for heat dissipation. 

       When phonons transport across a material interface, they will experience reflection, transmission, and mode conversion, meanwhile, these properties are highly interface morphology-, size-, and defects-dependent  \cite{li2012size} \cite{li2012effect}.To  clarify the underlying physics, the acoustic mismatch model (AMM) and diffuse mismatch model (DMM) are two early empirical models that are often used to describe the phonon transmission at the interface. Although some interfacial thermal conductance (ITC, reciprocal of ITR) values have been successfully predicted using these models while most deviate significantly from the experimental works due to their limit assumptions of phonons: complete specular scattering and complete diffuse scattering. Atomistic methods, such as atomistic Green’s function method and molecular dynamics (MD) method, can take into account the actual atomic structures of the interface, and based on these methods various realistic factors, such as atomic inter-diffusion  \cite{liu2023impact},interface roughness   \cite{zhou2013relationship} \cite{wang2023atomic}, and dislocations  \cite{li2020gan},have been demonstrated significantly affect the nanoscale thermal transport.  It should be noted the transmission coefficient is obtained with atomistic methods typically as a function of frequency, which describes a collective/averaged performance of multi phonon modes with different polarizations and buries the dynamic behavior of specific phonon mode at the interface. Actually, different phonon modes perform differently in interface thermal transport. For instance, very recently, Li et al.  \cite{li2022atomic} experimentally observed that distinct types of phonons at AlN/Si interface can mutual penetration forming phonon bridges across the interface, such as Si-LA/LO mode connect to AlN-TA mode, and Si-TO mode connects to AlN-TO mode. To gain insight into these deep knowledge, in this work, we resort to the phonon wave packet simulation, which can give intuitive observation of phonon scattering of specific mode and without any phonon scattering mechanism assumption, to explore the phonon dynamic behaviors crossing the heterointerface.

       In realistic material interfaces, a thin (few nm) amorphous interlayer due to lattice and thermal mismatch between dissimilar materials inevitably occurs in solid-solid interfaces  \cite{cheng2020interfacial} \cite{mu2019high}. Some studies reporting that the amorphous interlayer improves the ITC, for example, Tian et al.  \cite{tian2012enhancing} found that the amorphous interlayer can soften the abrupt change of acoustic impedance at Si/Ge interface and bridge the vibrational mismatch between Si and Ge. However, for the majority of heterostructures, such as GaN/diamond  \cite{cheng2020interfacial}, GaN/AlN  \cite{wang2021interfacial}, GaN/SiC  \cite{mu2019high}, etc, the amorphous interlayer was reported to reduce the ITC significantly since amorphous material thermal conductivities are typically very low by comparison to crystals. In addition, due to structurally disordered, only propagons exhibit a propagating nature in amorphous material \cite{gordiz2016phonon, gordiz2017phonon, zhou2020thermal}. Thus, the phonon localization phenomenon was apt to occur in disordered systems, such as Si/Ge heterostructure with interface atomic diffuse  \cite{ni2019interface}, GaAs superlattices with randomly distributed ErAs particles  \cite{mendoza2016anderson}. For these cases, the phonon participation ratio or the exponential decay features of phonon transmission coefficients typically as an indicator to indirectly characterize the phonon/Anderson localization, however, the direct observation of phonon localization in experiments is still challenging because of the spatial resolution limitation of conventional optical phonon detection techniques and extremely weak phonon signal from localization energy  \cite{zhang2017impeded} \cite{gadre2022nanoscale}. Although some interesting physical phenomena induced by the amorphous interlayer have been reported by theoretical simulation, such as reducing the phonon non-equilibrium near the interface  \cite{li2022inelastic}, there are still many ambiguous issues caused by the amorphous interlayer needed to clear. For example, whether the amorphous interlayer can alter the mode character after phonon-interface scattering? Whether the interfacial phonon dynamics depends on the size of the amorphous interlayer? How the amorphous interlayer affects all phonon modes’ contribution to the ITC? 

           To address these issues, in this work, we use GaN/AlN as a prototype system, which widely exists and plays a vital role in GaN-based power devices, to explore the dynamic behaviors of phonons at the interface with a thin amorphous interlayer by using phonon wave packet simulation. The thickness (\textit{L}) of amorphous interlayer is analogous to the experimental observation in GaN/AlN heterointerface that varied from 1$\sim$3 nm  \cite{spindlberger2021cross} \cite{zheng2019study}.A phonon wave packet is initialized with a predefined frequency and polarization and allowed to propagate through the interface. The dynamic behaviors of phonons at interface are monitored by analyzing the atomic displacements and velocities in real space and reciprocal space. The phonon transmissivity is calculated by measuring the phonon energy change before and after interface scattering. To  assess the effects of the amorphous interlayer on ITC, the ITCs of GaN/AlN with different interface morphologies and amorphous interlayer thickness are calculated by summation over all phonon contributions with the Landauer formula. Finally, to improve the phonon transmission, an optimized GaN/AlN interface morphology was developed via the annealing reconstruction technique, and the resulting phonon transmission and ITC were discussed. 

\section{Models and methods}

\subsection{Model introduction.}

Fig. 1(a) shows the three types of GaN/AlN heterointerfaces studied in this work, sharp interface, amorphous interface, and reconstructed interface. Sharp interface is an ideal interface, which cannot be realized due to the dislocation, stress, and high temperature growth condition. Amorphous interface was attained by locally melting the interface at 4000 K and quenched to 300 K using Nose$-$Hoover reservoirs. Reconstructed interface is an optimized interface, which was obtained by annealing and recrystallization of the amorphous interlayer, detailed introduction for this interface will be discussed subsequently. In our simulation, both GaN and AlN are wurtzite structures and aligned in [0001] (\textit{z}) direction, mimicking the experimental epitaxial condition that GaN grows on AlN \cite{stanchu2018strain}. The atomic interactions are described using the Stillinger-Weber (SW) potential with the parameters developed by Bere and Serra (for GaN) \cite{bere2002atomic} and Lei (for AlN) \cite{lei2009molecular}, and the interaction between GaN and AlN at interface are created by the mixing rule. Using these potentials, the phonon spectra for GaN and AlN along the high symmetry $\Gamma$-A ([0001]) direction were calculated, as shown in Fig. 1(b), via the harmonic lattice dynamics calculations. The unfilled circles correspond to the experimental data that are obtained by Ruf (for GaN) \cite{ruf2001phonon} and Schwoerer (for AlN) \cite{schwoerer1999phonons} with the inelastic X-ray scattering, which indicates that the potentials we used can accurately reproduced the essential vibration properties in GaN and AlN. Because GaN and AlN have the same atomic structure and only differ in atomic mass, thus, a similar dispersion curve is observed for both materials, except that the frequency in GaN is lower than that in AlN by a factor of\(\sqrt{m_{GaN}/m_{AlN}}\approx 1.4\). The lattice constants of GaN and AlN are \textit{a\textsubscript{GaN}} $=$ 3.18 Å, \textit{c\textsubscript{GaN}} $=$ 5.21 Å and \textit{a\textsubscript{AlN}} $=$ 3.09 Å, \textit{c\textsubscript{AlN}} $=$ 5.05 Å, which give rise to a negligible lattice mismatch strain, \textit{$\chi$} $=$ (\textit{a\textsubscript{GaN} - a\textsubscript{AlN}}) / \textit{a\textsubscript{GaN}} $=$ 2.8$\%$. 

\subsection{Phonon wave packet simulations.}

To investigate the dynamic behaviors of phonons at interface, the phonon wave packet simulations were performed using the LAMMPS package  \cite{plimpton1995fast}.A wave packet is formed from a linear combination of  plane waves \textit{$\varepsilon$} of a certain polarization \textit{v} and centered at a specified wave vector \textit{k}. The initialized atomic displacement \textit{u\textsubscript{l}\textsubscript{,b}} for the \textit{b}th atom in the \textit{l}th unit cell in a wave packet can be assigned by the following equations:

\begin{equation}
u_{l,b}=\frac{A}{\sqrt{m_{b}}}\varepsilon_{v,k_{0},b}(k_{0})exp[ik_{0}(z_{l}-z_{0})-\omega t]\times exp[-\frac{(z_{l}-z_{0})^{2}}{xi^{2}}]
\end{equation}

where \textit{A} is the wave packet amplitude,\textit{m\textsubscript{b}} is the atomic mass of the \textit{b}th atom, \textit{$\varepsilon$\textsubscript{v,k0,b}} is the eigenvector for the \textit{b}th atom in a unit cell with wave vector \textit{k\textsubscript{0}} and polarization \textit{v}, \textit{$\omega$} is the angular frequency of the phonon mode, \( z_{l}\) is the coordinate of the \textit{l}th unit cell, and \textit{$\xi$ }is the spatial extent of the wave packet. The initialized propagation velocity of the wave packet is created by the time derivative of the initialized atomic displacement. Using this equation, a Gaussian wave packet is generated centered in real space around \textit{z}\textsubscript{0} and in reciprocal space around \textit{k}\textsubscript{0}. Usually, a long model system is required in wave packet simulation, here the entire system length in the \textit{z}-direction was chose 1100 unit cells (600 unit cells for AlN, 500 unit cells for GaN). To avoid the influence of other modes, the whole system was first carefully relaxed to make the atoms around their equilibrium positions ($\sim$0 K). After that, one wave packet with designated phonon properties that are extracted from the phonon dispersion curve in Fig. 1(b) is launched from the AlN side. When the wave packet encounters the interface, scattering events take place, part of phonon energy can transmit through the interface, and some is reflected back. According to the fraction of phonon energy transmitted through the interface, the phonon transmission coefficient \textit{$\alpha$} can be determined as follows

\begin{equation}
\alpha  =\frac{E_{tran}}{E_{inci}}
\end{equation}

where \textit{E\textsubscript{tran}} is the transmitted phonon energy, \textit{E\textsubscript{inci}\textsubscript{ }}is the incident phonon energy. Further, using the calculated \textit{$\alpha$}, the mode-resolved ITC, G, can also be estimated by summation over all phonon contributions with the Landauer formula \cite{wei2019phonon}

\begin{equation}
G =\frac{1}{(2\pi )^{2}}\sum_{k}^{}\sum_{\lambda }^{}W_{k}\overline{h}\omega_{\lambda k}v_{z,\lambda k}\alpha_{\lambda k}\frac{dN(\omega ,T)}{dT}
\end{equation}

where \textit{W\textsubscript{k}} is the weighting factor associated to the \textit{k}-space volume, \(\overline{h}\) is the reduced Planck constant, \textit{v\textsubscript{z$\lambda$k}} is the phonon group velocity that determined by the displacement per unit time of the wave packet, \textit{$\alpha$\textsubscript{$\lambda$k}} is the phonon transmission coefficient that obtained from the phonon wave packet simulations, \textit{N}(\textit{$\omega$}, \textit{T}) is the Bose-Einstein distribution at temperature \textit{T}. 

Theoretically, the wave packets can be analyzed either from the real space or from the reciprocal space. Comparing to the real space, the reciprocal space can provide more information like the phonon frequency, mode splitting and conversion after interface scattering. Thus, by Fourier transform of atomic displacements, we obtained the normal coordinates ($\psi$) of specific wave packet after scattering in reciprocal space

\begin{equation}
\Psi(k,t)=\sqrt{\frac{m_{b}}{N}}\sum_{lb}u_{l,b}(t)\varepsilon_{b\lambda k}^{\ast}exp[-ik(z_{l}-z_{0}].
\end{equation}

\begin{figure}[!htbp]
\centering
\begin{subfigure}[b]{0.45\textwidth}
\centering
\includegraphics[width=\textwidth]{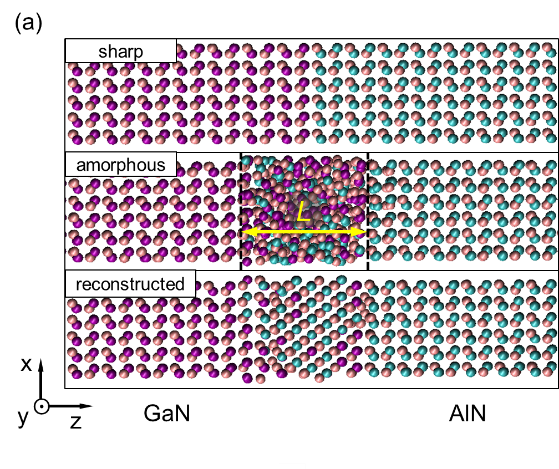}
\end{subfigure}
\hfill
\begin{subfigure}[b]{0.45\textwidth}
\centering
\includegraphics[width=\textwidth]{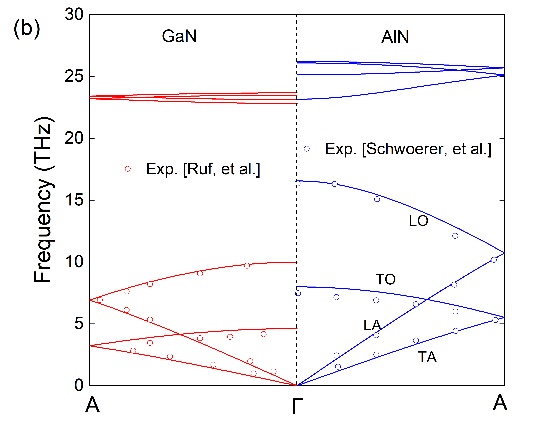}
\end{subfigure}
\end{figure}

\textbf{Figure. 1}.(a) Schematic of GaN/AlN heterostructures with sharp, amorphous, and reconstructed interfaces. (b) Phonon dispersion curves for GaN and AlN, respectively. The unfilled circles correspond to the experimental data that are obtained by Ruf (for GaN \cite{ruf2001phonon}) and Schwoerer (for AlN \cite{schwoerer1999phonons}) with inelastic x-ray scattering.

\section{Results and discussion}

\subsection{Dynamic behavior of phonon passing through GaN/AlN interface}

We first explored the dynamic behaviors of phonons at GaN/AlN interface with an amorphous interlayer, meanwhile, the phonon propagation at sharp interface also be displayed for comparison. As shown in Fig. 1(b), GaN and AlN have one LA branch and two degenerate TA branches. From the eigenvectors, we found that the TA mode is lateral (\textit{x-y} direction) transverse-polarized and perpendicular to the wave vector. In contrast, the LA mode is not strictly axial (\textit{z} direction) longitudinal-polarized and has a lateral square displacement (<0.02$\%$). This is due to the fact that the GaN and AlN both are not symmetric with respect to the (0001) plane (the \textit{x-y} plane in our simulation). In this work, the TA modes are represented by the lateral atomic displacements that are averaged in the x and y directions. 

Fig. 2 shows several representative snapshots of wave packets after scattering at the interface, the gray lines refer to the initialized atomic displacement in axial/lateral direction and the red lines refer to the excited atomic displacement in lateral/axial direction. For sharp interface, it is found that the energy of incident wave packet from AlN is primarily transformed into the mode with identical polarized in GaN. For example, in Fig. 2(a), a LA polarized wave packet is launched from AlN side with a frequency of 1.2 THz. After being scattered at the interface, about $\sim$97.6$\%$ of the incident energy transmits into GaN side with LA polarization, and the remaining energy is reflected and maintains the LA character. Actually, there has a small amount of phonon energy, less than 0.1$\%$ of the incident energy, is converted into TA polarization and across the interface, as shown in the inset of Fig. 2(a). This unobvious LA\( \rightarrow\)TA mode conversion behavior is related to the lateral components in the phonon eigenvector of LA mode and nearly plays a negligible role in interface thermal transport. However, when the interface with an amorphous interlayer, we found that this mode conversion behavior is remarkably enhanced. As shown in Fig. 2(b), when a same LA wave packet scatters at the amorphous interface, here \textit{L}$=$2 nm, the excited (transmitted and reflected) TA mode energy by LA\( \rightarrow\)TA mode conversion increases to the $\sim$4$\%$ of the incident energy. And with the frequency increases, the proportion of phonon energy used for mode conversion further increases. In addition, we found that the amorphous interlayer can also convert a portion of acoustic mode energy into optical mode, as shown in Fig. 2(c), where we launched a 3.6 THz LA wave packet. According to the propagate velocities of wave packets and the form of atom vibration (as shown in the inset of Fig. 2(c), the adjacent-layer atoms move ``in-phase" referring to acoustic phonons and different types of atoms move ``out-of-phase" referring to optical phonons), we verified that the reflected wave packets include two acoustic modes (LA + TA) and the transmitted wave packets include one acoustic mode (LA) and one optical mode (TO). This can be explained by the phonon spectra, where two low-frequency optical branches follow closely to the LA and TA branches. Clearly, the frequency 3.6 THz of incident wave packet just falls into the range of low-frequency TO (3.2 $\sim$ 4.6 THz) of GaN. Therefore, the mode conversion from acoustic phonons to optical phonons is possible. 

Although such mode conversion behavior has not been observed in TA (no axial component in the eigenvector) mode phonon incident at the sharp interface, it occurs at amorphous interface. As shown in Fig. 2(d), we provided one example with TA\( \rightarrow\)LA mode conversion, where a 1.14 THz TA mode wave packet excited two LA modes after interface scattering. Different from the LA modes, with frequency increases, we surprisingly found a portion of high-frequency TA mode phonons that cannot transmit in sharp interface can transmit due to the existence of amorphous interlayer. Even the frequency is larger than the cut-off frequency, $\sim$4.6 THz, of TO mode in GaN. As shown in Fig. 2(f), a TA mode phonon wave packet with a frequency of 4.8 THz scattered at amorphous interface, it is obvious there is still a considerable fraction of phonon energy that can be transmitted. Generally, when the frequency of incident wave packet beyond the cut-off frequency on the other side would be completely reflected because lacking a corresponding thermal transport channel \cite{liu2020analytical}. As shown in Fig. 2(e), this same TA mode wave packet is reflected by the sharp interface completely. Thus, there are other phonon transmission channels were created at amorphous interface: TA\( \rightarrow\)LA mode conversion and diffuse scattering. Meanwhile, this finding points out another reason why the AMM model cannot describe accurate phonon transmission coefficients, which neglects the mode conversion \cite{liu2020analytical}. 

\begin{figure}[!htbp]
	\centering
	\includegraphics[width=11.9cm,height=9.95cm]{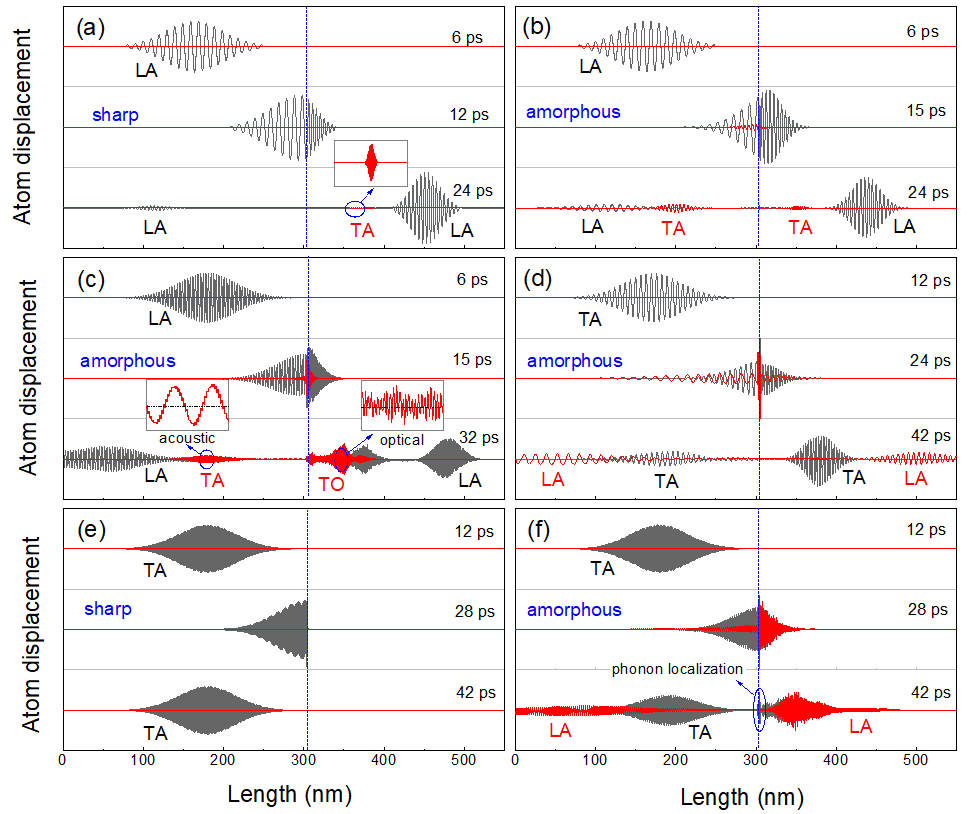}
	\caption{Snapshots of wave packets crossing the GaN/AlN interface with different polarizations, frequencies, and interfaces. The gray lines refer to the initially atom displacements, the red lines refer to the excited atom displacements after phonon-interface scattering. (a) LA, $\omega$=1.2 THz, sharp interface. (b) LA,$\omega$=1.2 THz, amorphous interface. (c) LA, $\omega$=3.6 THz, amorphous interface (L=2nm). (d) TA, $\omega$=1.14 THz, amorphous interface. (e) TA, $\omega$=4.8 THz, sharp interface. (f) TA, $\omega$=4.8 THz, amorphous interface (L=2nm). The inset in Fig. 4(a) shows the excited TA polarized phonon wave packet. The inset in Fig. 4(c) shows the form of atom vibration in different wave packets. The inset in Fig. 4(f) refers to the localized phonon around the interface.}
	\label{fig:fig2}
\end{figure}

Besides the contribution from mode conversion and diffuse scattering, inelastic scattering probably occur as phonons pass through the amorphous interlayer, resulting in part high-frequency phonons being scattered into multiple lower frequency phonons and then transmitted. 

To verify it, we analyzed the frequency of wave packets after scattering at sharp and amorphous interfaces, respectively, by Fourier transform of atom displacement. As shown in Fig. 3, we choose three TA mode wave packets with different frequencies of 1.14 THz, 2.28 THz, and 4.1 THz. The central location for each peak corresponds to the frequency of wave packet, and the magnitude of peak equals to the amplitude and scales quadratically (\textit{E}(\textit{k})$\propto$\textit{$\psi$}\textsuperscript{2}(\textit{k})) with the energy of wave packet. The transmitted and reflected wave packets are represented by the dotted lines. For sharp interface, it is found that the transmitted and reflected wave packets have almost the same frequency as the original incident wave packets, indicating the phonon transport process at sharp interface is completely elastic. However, when the interface with an amorphous interlayer, an obvious frequency shift occurred especially for the transmitted wave packets. For example, for an incident TA wave packet with 1.14 THz, the frequency of the transmitted TA wave packet shifts to $\sim$1 THz. As we all known, the elastic process does not involve frequency conversion, thus, the frequency shift here results from the inelastic scattering. Similar inelastic scattering behaviors with wave packet simulation have also been observed in Si/Ge \cite{deng2014kapitza} and Si/SiO\textsubscript{2} \cite{sun2010molecular} interfaces. It should be noted that, as the incident frequency increases, such as at 4.1 THz, either reflected or transmitted wave packet was found to span a broader frequency or wave vector range. This can be understood by the superposition principle of waves. Theoretically, one wave packet is a linear combination of plane waves that with a certain polarization and centered at a specified wave vector. Hence, the broadening of frequency means additional phonon modes that result from inelastic scattering are participating in building wave packets. 

\begin{figure}
	\centering
	\includegraphics[width=7.36cm,height=6.14cm]{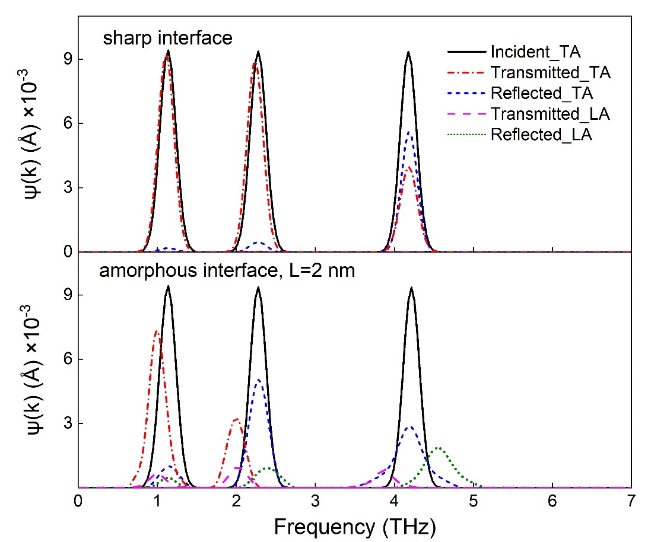}
	\caption{The normal coordinates of wave packets in reciprocal space after the interface scattering. The black lines refer to the incident TA modes, the dotted lines refer to transmitted and reflected phonon modes. }
	\label{fig:fig3}
\end{figure}

\subsection{Interfacial phonon transmission}

By recording the transmitted phonon energy, we further assessed the effects of the amorphous interlayer on phonon transmission. As shown in Fig. 4(a-b), the phonon transmission coefficients of LA and TA modes as a function of frequency and amorphous interlayer thickness (\textit{L }$=$ 1 nm, 2 nm, and 3 nm). For sharp interface, the phonon transmission coefficients show a smooth decrease with an increase in frequency. In particular, transmission coefficient changes continuously even at the cut-off frequency of LA and TA branches of GaN, presumably because of the continuity in the LA-LO and TA-TO dispersion curves. As expected from the analysis of the dispersion curves, the transmission coefficients of LA and TA attenuate to zero at $\sim$ 9.5 THz and $\sim$ 4.6 THz, respectively. However, when the interface has an amorphous interlayer, the phonon transmission coefficient of LA and TA fluctuates and declines sharply with the increase of frequency because of the enhanced phonon Umklapp scattering. And with thickness \textit{L} increases, phonon scattering will be occurred in a broader spatial region, which results in a shorter phonon mean free path, hence the phonon transmission shows a further decrease. According to equation (3), the reduction of phonon transmission will lead to a negative effect on interface thermal transport. 

For TA mode, as mentioned above we found that the amorphous interlayer leads to some unexpected scattering outcomes, where some phonons cannot transmit in sharp interface can transmit. This anomalous finding can also be confirmed by the TA mode phonon transmission in Fig. 4(b), in which the transmission coefficient of high-frequency (>4 THz) TA phonons exceeds the sharp interface. Meanwhile, in this high-frequency range, it is noted that the phonon transmission curves flatten out and gradually became insensitive with the frequency. This phenomenon can be interpreted by the specular-diffuse scattering model  \cite{shao2018understanding},when the phonon wavelength is much  smaller than the thickness of amorphous interfacial layer, the fraction of phonon specular scattering at interface will dramatically decrease even vanish. Therefore, the flattening of the trend is largely attributed to the full-diffusion effect, which results in the memory of the incident phonon lost. In Fig. 4(b), we observed that the transmission coefficient of TA phonon ultimately converges to 0.3$\sim$0.5, which approaches the DMM that predicts $\sim$50$\%$ transmission  \cite{majumdar2004role}, thus is a typical phonon diffusion result.

\begin{figure}[!htbp]
	\centering
	\includegraphics[width=12.87cm,height=9.68cm]{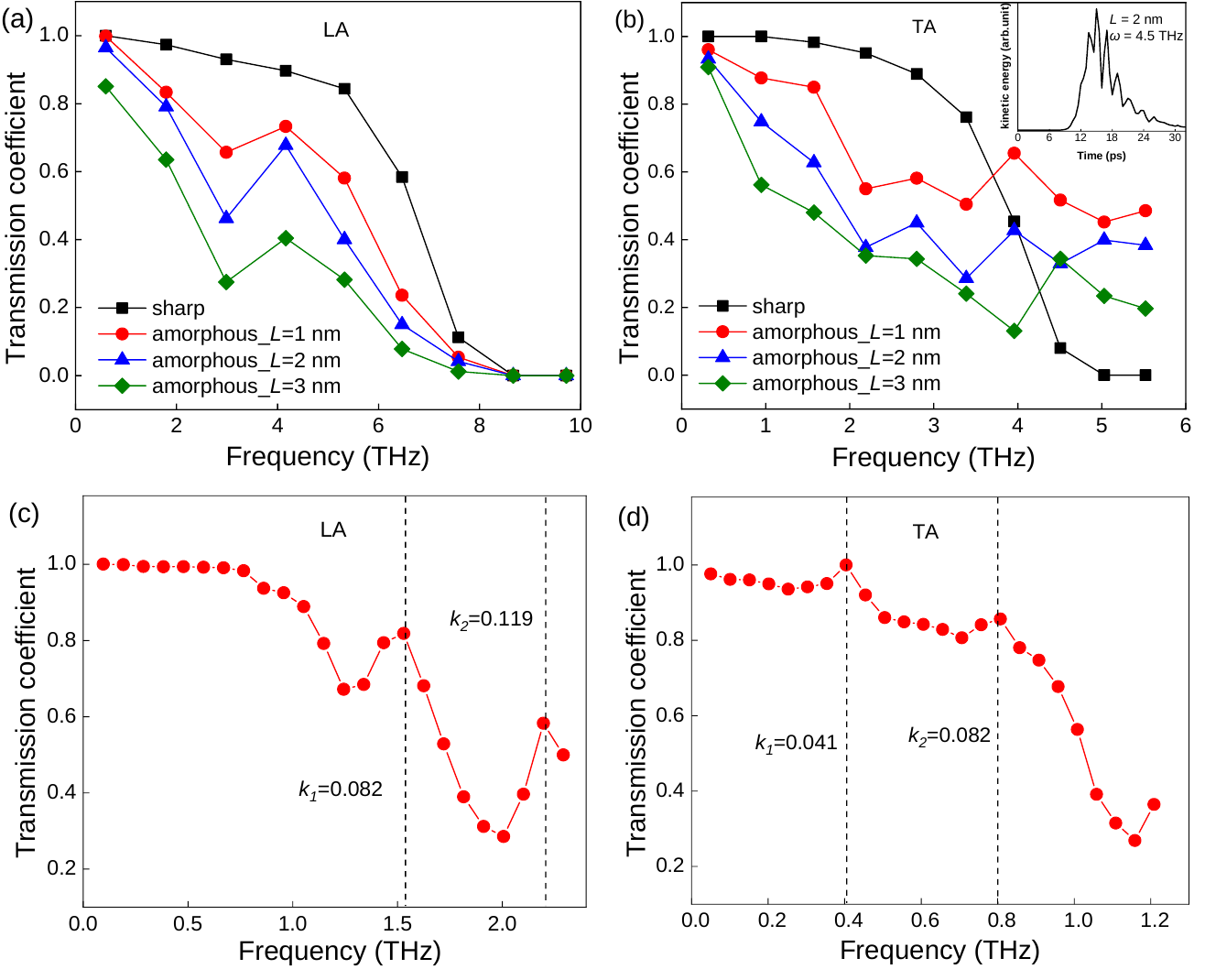}
	\caption{The energy transmission coefficients as a function of phonon frequency. (a) LA and (b) TA mode energy transmission coefficients for different interface morphologies and amorphous layer thickness. The low-frequency (a) LA and (b) TA energy transmission coefficients at amorphous interface with L=8 nm. The inset in (b) shows the kinetic evolution in the amorphous layer (L=2nm) when the wave packet crosses the interface.}
	\label{fig:fig4}
\end{figure}

\subsection{Phonon localization in amorphous interlayer}

Another noteworthy thing is that the convergent TA mode transmission coefficient in the high-frequency region decreases with the increase of the amorphous interlayer thickness. Apart from the increased Umklapp scattering, phonon localization was revealed to be another significant impact factor. As shown in Fig. 2(f), an evident spatially localized phonon energy was trapped in the amorphous interlayer, and the fraction of the localized energy was found to increase with the increase of the thickness of the amorphous interlayer. Similar energy localization has also been reported previously in composition graded Si/Ge  \cite{zhang2017impeded} interface and disordered Si  \cite{shao2017probing} surface. Their studies suggested that the fluctuation of wave speed caused by the non-uniform mass density of the amorphous layer is an important factor of phonon localization. Theoretically, when the wave packet enters the amorphous interlayer, it cannot be supported by the disordered structure and will excite some random vibrational modes in amorphous interlayer, which are called locons or localized resonant modes  \cite{zhou2020thermal}.Thus, part of incident wave packet  energy will be converted into these random vibrational modes, and then be slowly emitted in the form of a high-frequency lattice ripple. This process can be confirmed by monitoring the kinetic energy evolution in the amorphous interlayer, as shown in the inset of Fig. 4(b), where we recorded the kinetic energy variation in the amorphous interlayer (\textit{L}$=$2 nm) before and after a TA mode (\textit{$\omega$}$=$4.5 THz) wave packet passed through the interface. We find that the kinetic energy in the amorphous interlayer first increases rapidly as the wave packet enters the amorphous interlayer and then decays gradually through a re-emission process. Even through the wave packet crossing and moving away from the interface, we found there is still about $\sim$0.46$\%$ of total incident energy holding inside the amorphous interlayer in the form of localized phonons or quasi-standing waves. Since localization virtually ``immobilizes" phonons and makes them non-conducting, namely, these phonons lose their nature as heat carriers. Hence, we conclude that the phonon localization in the amorphous interlayer will aggravate heat accumulation and further decay the efficiency of energy dissipation in devices.  

\subsection{Phonon interference induced by the amorphous interlayer}

As aforementioned, the phonon transmission coefficient fluctuates and shows an apparent oscillation behavior when the wave packets pass through the amorphous interlayer, as shown in Fig. 4(a-b). This is likely due to an interference effect. Since the amorphous interlayer actually generates two interfaces: AlN/interlayer interface and interlayer/GaN interface, which lead to the multiple scattering of incident wave packet. Specifically, when an incident wave packet hits on the AlN/interlayer interface, a portion of wave packet energy is transmitted while the other is reflected. The transmitted wave packet soon reaches the interlayer/GaN interface, producing transmitted and reflected wave packets again. These two reflected wave packets may overlap and therefore induce interference. To verify if the oscillatory behavior of transmission coefficient is caused by phonon interference, we compare the transmission coefficients from wave packet simulation with the values from the two-beam interference analytical equation  \cite{liang2017phonon}, 

\begin{equation}
I = I_{1}+I_{2}+2\sqrt{I_{1}I_{2}}cos(2kL+\varphi_{0}),
\end{equation}

where \textit{I} is the intensity of the interference signal, i.e., the transmitted wave packet energy in our simulations; \textit{I\textsubscript{1}} and \textit{I\textsubscript{2}} are reflections at the two interfaces, respectively; \( \varphi_{0}\) is initial phase of the interference, \textit{L} is the thickness of the amorphous interlayer, and \textit{k} is the wave vector of wave packet. Because the two adjacent interference maxima have a phase difference of 2\textit{$\pi$}, thus, the interference period \textit{t\textsubscript{p}} can be defined as 

\begin{equation}
t_{p} = k_{v,n+1}-k_{v,n} =\frac{\pi}{L},
\end{equation}

where \( k_{v,n+1}\) and \( k_{v, n}\) are the center wave vectors of two adjacent peaks in the interference spectrum. Obviously, the interference period \textit{t\textsubscript{p}} is inversely proportional to the thickness of the amorphous interlayer. Considering high-frequency phonon with high possibility of diffuse scattering, which can lead to decoherence of phonons and then make the observation of wave behavior challenging. Thus, to have a better understanding, we enlarged the thickness of amorphous interlayer to 8 nm, and recalculated the phonon transmission coefficient with a wave vector range from 0.01 $\times$ (2$\pi$/\textit{c}\textsubscript{AlN}) to 0.2 $\times$ (2$\pi$/\textit{c}\textsubscript{AlN}). Correspondingly, by substituting \textit{L}$=$8 nm into equation (6), the interference period with theoretical prediction is 0.39 nm\textsuperscript{-1}.

Fig. 4(c-d) shows the calculated phonon transmission coefficients at this low-frequency range, it is clear that the transmission coefficient of both LA and TA modes exhibits an evident oscillatory behavior. According to the center frequencies (\textit{$\omega$}\textsubscript{1},\textit{ $\omega$}\textsubscript{2}) of the two adjacent peaks in Fig. 4(c-d), we obtained their corresponding wavevectors (\textit{k}\textsubscript{1}, \textit{k}\textsubscript{2}) with the help of phonon dispersion curves in Fig. 1(b). Here, for LA mode, \textit{k}\textsubscript{1}$=$0.82 nm\textsuperscript{-1} and \textit{k\textsubscript{2}}$=$1.19 nm\textsuperscript{-1}, similarly, for TA mode, \textit{k}\textsubscript{1}$=$0.41 nm\textsuperscript{-1 }and \textit{k}\textsubscript{2}$=$0.82 nm\textsuperscript{-1}. Further, the oscillatory periods for LA and TA are determined as 0.37 nm\textsuperscript{-1} and 0.41 nm\textsuperscript{-1}, respectively, which agree well with the interference period 0.39 nm\textsuperscript{-1 }in theoretical predication. Therefore, we conclude that the oscillatory behavior of transmission coefficient was ascribed to the phonon interference effect and the maxima and minima of transmission coefficient are the result of destructive and constructive interferences, respectively. 

\subsection{Interfacial thermal conductance}

To further quantify the phonon contributions at different interface morphologies, the mode-resolved ITCs were obtained with the calculated phonon transmission coefficients and Landauer formula. Strictly speaking, the cumulative ITC should be a summation of phonon contribution over the entire first Brillouin zone  \cite{gordiz2017phonon} \cite{wei2019phonon}. Here we simplified the calculation and quantified the phonons contribution perpendicular to the interface direction ($\Gamma$ - A direction), which has been demonstrated to be the major channel in GaN/AlN interfacial thermal transport  \cite{wang2007atomistic} \cite{ju2016anisotropic}. Specifically,  the Brillouin zone was discretized with a 40  40  40 \textit{k}-point grid, in which the 40 \textit{k}-points along the $\Gamma$ - A direction were used. Fig. (5) shows the ITC contributions from each phonon branch by equation (3) for sharp as well as amorphous interfaces. For sharp interface, LA and TA modes contribute $\sim$50$\%$ of the ITC across the interface, respectively. Compared with the sharp interface, the ITC with an amorphous interlayer shows a pronounced reduction, and with the increase of \textit{L} nearly monotonically decreases. When \textit{L}$=$3 nm, the ITC decreased by $\sim$54$\%$ from 176 MWm\textsuperscript{-2}K\textsuperscript{-1} to 82 MWm\textsuperscript{-2}K\textsuperscript{-1}. The decreased trend and magnitude in ITC are comparable to recent reports [9], in which the GaN/AlN interface with a compositional diffusion interlayer.   

\begin{figure}[!htbp]
	\centering
	\includegraphics[width=6.47cm,height=5.39cm]{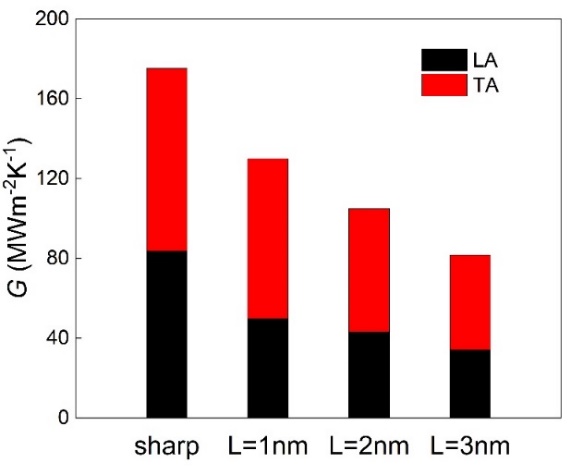}
	\caption{ Mode-resolved ITCs for sharp and amorphous (L = 1, 2, and 3 nm) GaN/AlN interfaces. }
	\label{fig:fig3}
\end{figure}

\subsection{Interface morphology optimization}

Given the remarkable ITR induced by the amorphous interlayer, we further optimized the interface morphology via annealing reconstruction technique. Recently, the effectiveness of this method has been experimentally demonstrated in heterogenous GaN/SiC interface  \cite{mu2019high}.Fig. 1(a) shows the reconstructed GaN/AlN interface, in which we found that the amorphous interlayer converted into  rocksalt phase instead of the initialized wurtzite phase after annealing. This finding has also been demonstrated in first-principle calculation  \cite{shulumba2016impact} and experimental observation  \cite{pravica2013high}, resulting from the state of the AlN rocksalt phase being more stable and energy being more favorable in comparison to its wurtzite phase. Taking \textit{L}$=$2 nm as an example, we calculated the phonon transmission coefficient for reconstructed interface using the wave packet simulations, as shown in Fig.  \cite{lee2016nanostructures}.And the amorphous interfac e with the same interlayer thickness also is displayed for comparison. Compared to the amorphous interface, the phonon transmission coefficients of the reconstructed interface at low- and mid-frequency regions have a remarkable improvement, indicating the reconstructed interlayer reduced the phonon scattering and then lead to a relatively large amount of phonon energy across the interface. Further, we also quantified the phonon contribution to ITC, as shown in Fig. 6(c), it is found that the ITC improved by $\sim$21$\%$ after annealing treatment. It should be noted that compared with the sharp interface, there still exists a large gap. This difference mainly attributed to two factors. First, the intrinsic thermal conductivity of rocksalt AlN is much lower than that of the wurtzite phase. Using ab initio simulation, Shulumba \textit{et al}.  \cite{shulumba2016impact} reported that the thermal conductivity of the rocksalt phase (81 Wm\textsuperscript{$-$1}K\textsuperscript{$-$1}) of AlN was only a quarter of the wurtzite phase (320 Wm\textsuperscript{$-$1}K\textsuperscript{$-$1}) due to the influence of phonon inelastic scattering. Second, the rocksalt phase altered the original crystallographic orientation, which can also cause additional phonon scattering and then suppress the phonon propagation forward. 

\begin{figure}[!htbp]
	\centering
	\includegraphics[width=12.82cm,height=10.6cm]{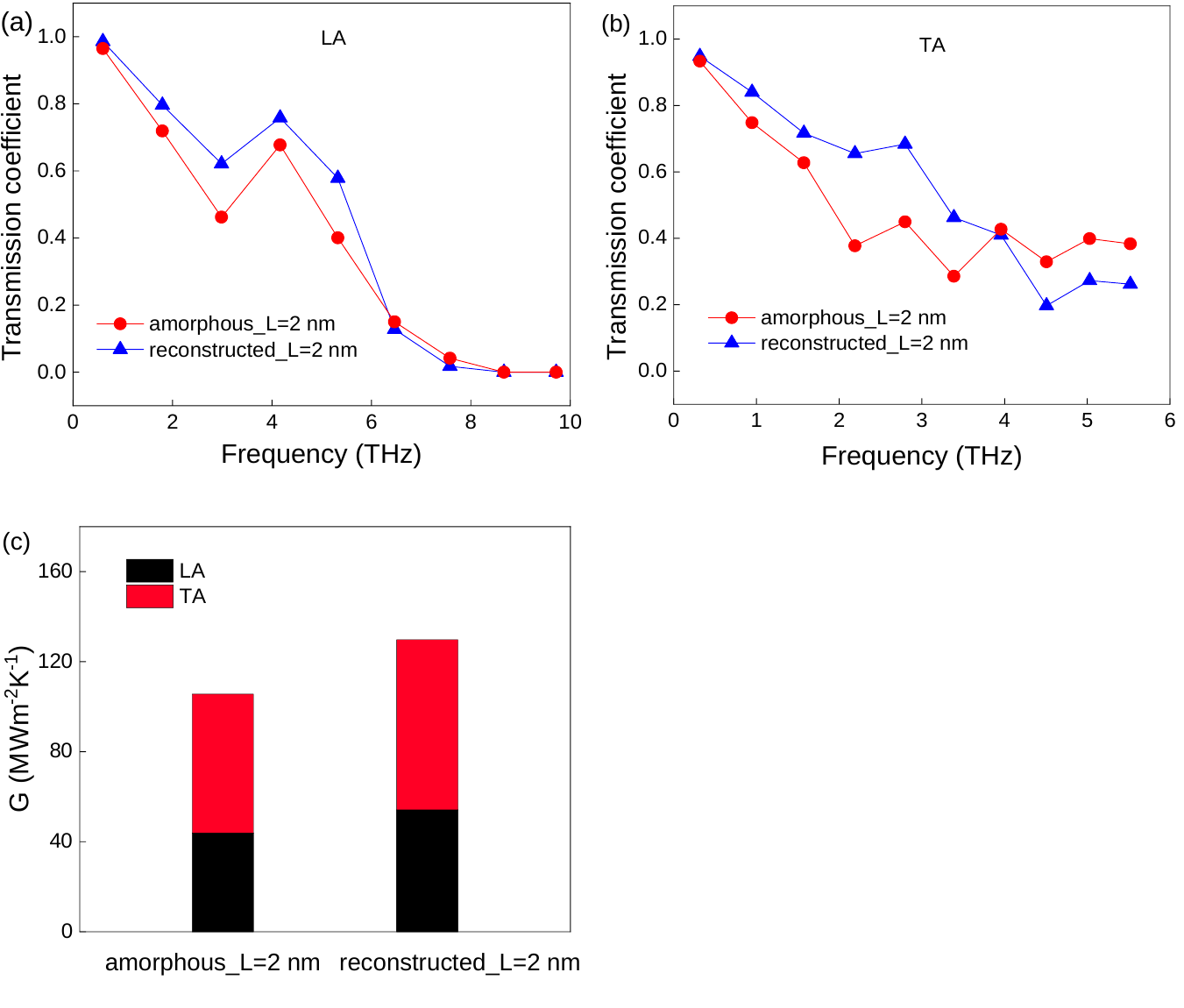}
	\caption{(a) LA and (b) TA mode energy transmission coefficients for amorphous and reconstructed interfaces. (c) Mode-resolved ITCs for amorphous and reconstructed interfaces.}
	\label{fig:fig4}
\end{figure}

\section{Conclusions}

In this work, using phonon wave packet simulation, we studied the phonon dynamic behaviors at GaN/AlN interface with an amorphous interlayer. We found that the amorphous interlayer impedes the phonon transport, and the resulting phonon transmission coefficient exhibits an obvious frequency and amorphous layer thickness dependency. Compared to the sharp interface, complicated phonon scattering events, such as mode conversion, frequency shift (or inelastic scattering), phonon interference, and phonon localization, were occurred when phonons pass through the interface. Due to the multiple scattering of phonons in the amorphous interlayer, the phonon transmission coefficient exhibits an oscillation behavior, and the oscillation period is consistent with the theoretical prediction using the two-beam interference equation. In addition, it is interesting to find that the high-frequency TA phonons transmissivity in the amorphous interface is abnormally higher than that across the sharp interface due to the mode conversion and inelastic scattering, which facilitate the redistribution of phonon energy. Finally, to aid phonon transmission, we further optimized the interface morphology by annealing reconstruction technique, which leads to an obvious improvement of the phonon transmission coefficients in low- and mid-frequency regions. Correspondingly, the ITC increases by 21$\%$ when \textit{L}$=$2 nm. Our study provides a deep understanding of the dynamic behaviors of phonons across the heterointerface and gives important insight into the physical origin of ITR. 

\textbf{ACKNOWLEDGMENTS}

The authors gratefully acknowledge the financial support from National Natural Science Foundation of China (Grant No. 52150610495 and Grant No. 52206080), the Shanghai Committee of Science and Technology (Grant No. 21TS1401500), and the Shanghai Municipal Natural Science Foundation (Grant No. 22YF1400100).

\bibliographystyle{unsrt}
\bibliography{references}

\end{document}